\documentclass[12pt,preprint]{aastex}
\usepackage{graphicx}
\begin{document}
\title{Strong gravitational lensing and galactic bulges}
\author{Da-Ming Chen}
\affil{National Astronomical Observatories, Chinese Academy of
Sciences, Beijing 100012, China}
\begin{abstract}
Lensing probabilities of quasars with image separations greater than
$\Delta\theta$ and flux density ratios less than $q_{\mathrm{r}}$
are calculated by foreground dark
matter halos in a flat, cosmological constant dominated
($\Lambda$CDM) universe . The mass density of the lenses is taken to be the
Navarro-Frenk-White (NFW) profile on all mass scales, plus a
central point mass for low-mass halos with
$M<M_c=5\times 10^{13}h^{-1}M_{\odot}$.
We introduce a quantity $M_{\mathrm{eff}}$, which is a point mass
ranging from $1$ to $1000$ times the mass $M_{\bullet}$ of a
supermassive black hole (SMBH) inhabiting the center of each
galaxy, to describe the contributions of galactic central SMBHs
and galactic bulges to lensing probabilities. The lensing cross
section and thus the lensing probability are quite sensitive to the
flux density ratio $q_{\mathrm{r}}$ of multiple images in our
calculations. It is shown that, to reproduce the lensing survey
results of JVAS/CLASS for $q_{\mathrm{r}}<10$, about $20\%$ of the
bulge mass is needed as a point mass for each galaxy. Since there
is still considerable uncertainty regarding the value of the
spectrum normalization parameter $\sigma_8$, we investigate the
effect of varying this parameter within its entire observational range
(from 0.7 to 1.1), and find that low $\sigma_8$ values ($\leq 0.7$) are
ruled out, and the best fit value is $\sigma_8\simeq 1.0$.
\end{abstract}
\section{Introduction}
Gravitational lensing provides us a powerful probe of the mass
distributions of the universe. By comparing the lensing probabilities
predicted by various cosmological models and the density profiles of
lenses with observations, we are able to test the mass distribution
of dark matter (CDM) halos and in particular,
the inner density slope in the sense that the JVAS/CLASS radio survey
\citep{browne,helbig,browne02,myers} has provided us the observed
lensing probabilities at small image separations
($0.3''<\Delta\theta<3''$).

It is well known that, Cold Dark Matter (CDM) model has become the
standard theory of cosmological structure formation. The
$\Lambda$CDM variant of CDM with
$\Omega_m=1-\Omega_{\Lambda}\approx 0.3$ appears to be in good
agreement with the available data on large scales \citep{primack}.
On smaller (sub-galactic) scales, however, there seems to be
various discrepancies. Issues that have arisen on smaller scales
have prompted people to propose a wide variety of alternatives to
the standard CDM model, such as warm dark matter (WDM) and
self-interacting dark
matter (SIDM). Now that problems arise from galaxy-size halos and
the centers of all dark matter halos, high-resolution simulations and
observations are the final criterion. Recent highest-resolution
simulations appear to be consistent with NFW \citep{klypin,power}
down to scales smaller than about 1 kpc. Meanwhile, a large set
of high-resolution optical rotation curves has recently been
analyzed for low surface brightness (LSB) galaxies, which suggests
that the NFW profile is a good fit down to about 1 kpc.
Although further simulations and observations, including the
observations of CO rotation curves \citep{bolat}, may help to
clarify the issue, it is likely that both WDM and
SIDM are probably ruled out, while the
predictions of $\Lambda$CDM at small scales may be
in better agreement with the latest data.

Unfortunately, if dark halos are modelled with NFW profile on all
mass scales, they will produce too few small angular separation
images in $\Lambda$CDM model as compared with the results of
JVAS/CLASS \citep{li}. A possible solution to the problem
is to modify the inner structure of dark halos by introducing
baryonic cooling and compression
\citep{porci,kocha,keeton2001a,sarbu,li,oguri}. Namely,
there exist two types of dark halos: small mass
halos (galactic size) with a steep inner density slope (singular
isothermal sphere, SIS) and very massive halos (e.g. galaxy clusters)
with a shallow inner slope (NFW). This can reproduce the observational
data of JVAS/CLASS \citep{li,sarbu}. One may also consider a unified
model in which dark halos are composed of an NFW-like component and
a bulge component for galaxies as lenses\citep{chen}. If the NFW
profile is believed to be universal, this model allow us to constrain the
structure of the galactic centers using the strong lensing surveys.

In this letter, we investigate the plausibility of the NFW+Bulge
model by fitting the observational data of JVAS/CLASS in a much
more accurate way to improve our previous work \citep{chen}.
Furthermore, we emphasize the importance of the flux density ratio
$q_{\mathrm{r}}$ of the two images produced by a central point
mass in each galaxy for the predicted results.

In our model, there are two important issues when galactic bulges are
involved: First,
the presence of supermassive black holes (SMBHs) at the
centers of most galaxies appears by now firmly established (Melia
\& Falcke 2001, and references therein). SMBH masses are estimated
to be in the range $10^6 - 10^9M_{\sun}$, and are
correlated with the masses and luminosities of the host spheroids and,
more tightly, with the stellar velocity dispersions
\citep{magor,ferra,ravin,merria,merrib,wandel,sarzi}. Recent
high-resolution observational data give
$M_{\bullet}/M_\mathrm{bulge}\approx 10^{-3}$\citep{merric}.
Furthermore, \citet{ferra02} finds a relation between masses
$M_{\bullet}$ of SMBHs and the total gravitational masses of the
dark matter halos as
$(M_{\bullet}/10^8M_{\sun})\sim
0.046(M_\mathrm{DM}/10^{12}M_{\sun})^{1.57}$.
Since these correlations extend well beyond the direct dynamical
influence of the SMBH it seems likely that there is a close link
between the formation of both the SMBH and its galaxy
\citep{silk,adams,haehn,islam,madau,menou,schne2002}. So we can
simply add galactic SMBHs as point masses to the NFW density profile
when we calculate lensing probability.

Second, galactic bulges, in which multiple
black holes may form and inhabit (e.g., Haehnelt \& Kauffmann
2002), can also contribute to the lensing probabilities at small image
separations. Since light rays are affected  by the mass within the
sphere of their impact distances, we can attribute the light
deflections induced by such a mass to an effective point mass,
which is referred to as $M_{\mathrm{eff}}$ in this paper. We thus
avoid the complexity of gas compression and mass distributions of
galactic bulges. Of course, we are unable to reveal the detail
structures of
the inner part of galaxies at the same time. We argue that
this may be an adequate way if we want to compare our
predictions with the results of strong gravitational lensing survey.
So, the total central pointlike masses by including
central SMBHs and bulges range from $1 - 1000M_{\bullet}$.
The upper limit of
$M_{\mathrm{eff}}$ is $1000M_{\bullet}$ because the bulge mass
correlates linearly with SMBH mass as
$M_{\bullet}/M_\mathrm{bulge}\approx 10^{-3}$.

\section{Lensing equation and probabilities}
We choose the most generally accepted values of the parameters for
the $\Lambda$CDM cosmology, in which, with usual symbols, the
matter density parameter, vacuum energy density parameter and
Hubble constant are respectively: $\Omega_{\mathrm m}=0.3$,
$\Omega_{\Lambda}=0.7$, $h=0.75$. The NFW density profile is
$\rho_\mathrm{NFW}=\rho_\mathrm{s}r_\mathrm{s}^3/
[r(r+r_\mathrm{s})^2]$, where $\rho_\mathrm{s}$ and $r_\mathrm{s}$
are constants. We can define the mass of a halo to be the mass
within the virial radius of the halo $r_\mathrm{ vir}$ :
$M_\mathrm{DM}=4\pi\rho_\mathrm{s}r_\mathrm{s}^3f(c_1)$, where
$f(c_1)=\ln(1+c_1)-c_1/(1+c_1)$, and $c_1=r_\mathrm{
vir}/r_\mathrm{s}=9(1+z)^{-1}(M/1.5\times
10^{13}h^{-1}M_{\sun})^{-1}$ is the concentration parameter, for
which we have used the fitting formula given by \citet{bul01}.

The surface mass density for a halo  as lens  is
\begin{equation}
\Sigma(\vec{x})=M_{\mathrm{eff}}\delta^2(\vec{x})
+\Sigma_\mathrm{NFW}(\vec{x}),
\end{equation}
where $x=|\vec{x}|$ and $\vec{x}=\vec{\xi}/r_\mathrm{s}$,
$\vec{\xi}$ is the position vector in the lens plane.
$\delta^2(\vec{x})$ is the two dimensional Dirac-delta function,
and $\Sigma_\mathrm{NFW}(\vec{x})$ is the surface mass density for
an NFW profile. $M_{\mathrm{eff}}$ is a point mass ranging from $1$
to $1000$ times the mass $M_{\bullet}$ of a supermassive black
hole (SMBH) inhabiting the center of each galaxy, to describe the
contributions of galactic central SMBHs and galactic bulges to
lensing probabilities. For galaxy cluster, $M_{\mathrm{eff}}=0$.
The lensing equation for this model is then
\begin{equation}
y=x-\mu_\mathrm{s}\frac{f_{\mathrm{eff}}+g(x)}{x},\label{lq}
\end{equation}
where $y=|\vec{y}|$,
$\vec{\eta}=\vec{y}D_\mathrm{S}^\mathrm{A}/D^\mathrm{A}_\mathrm{L}$
is the position vector in the source plane, in which
$D_\mathrm{S}^\mathrm{A}$ and $D_\mathrm{L}^\mathrm{A}$ are
angular-diameter distances from the observer to the source and to
the lens, respectively. It should be pointed out that, since the
surface mass density is circularly symmetric, we can extend both
$x$ and $y$ to their opposite values in Eq.(\ref{lq}) for
convenience. The parameter
$\mu_\mathrm{s}=4\rho_\mathrm{s}r_\mathrm{s}/\Sigma_\mathrm{cr}$
is independent of $x$, in which $\Sigma_\mathrm{cr}$ is critical
surface mass density. The term $f_{\mathrm{eff}}=2.78\times
10^{-4}f(c_1)M_{15}^{0.57}(M_{\mathrm{eff}}/M_{\bullet})$, where
$M_{15}$ is the reduced mass of an NFW halo defined as
$M_{15}=M_\mathrm{DM}/(10^{15}\mathrm{h}^{-1}M_{\sun})$, stands
for the contribution of a point mass $M_{\mathrm{eff}}$, and, of
course, $f_{\mathrm{eff}}=0$ for cluster-size lenses. The function
$g(x)$ stands for the contribution of the NFW halo, and it has an
analytical expression originally given by \citet{barte}.

When the quasars at redshift $z_{\mathrm{s}}=1.5$ are lensed by
foreground CDM halos of galaxies and clusters of galaxies, the lensing
probability with image separations larger than $\Delta\theta$ and
flux density ratio less than $q_{\mathrm{r}}$ is \citep{schne}
\begin{equation}
P(>\Delta\theta, <q_{\mathrm{r}})=
\int^{z_{\mathrm{s}}}_0\frac{dD_{\mathrm{L}}(z)}
{dz}dz\int^{\infty}_0\bar{n}(M,z)\sigma(M,z)B(M,z)dM,
\label{prob1}
\end{equation}
where $D_{\mathrm{L}}(z)$ is the proper distance from the observer
to the lens located at redshift $z$. The physical number density
$\bar{n}(M,z)$ of virialized dark halos of masses between $M$ and
$M+dM$ is related to the comoving number density $n(M,z)$ by
$\bar{n}(M,z)=n(M,z)(1+z)^3$, the latter is originally given by
\citet{press74}, and the improved version is
$n(M,z)dM=(\rho_0/M)f(M,z)dM$, where $\rho_0$ is the current mean
mass density of the universe, and
$f=(0.301/M)(d\ln\Delta_{\mathrm{z}}/d\ln
M)\exp(-|\ln(\Delta_{\mathrm{z}}/1.68)+0.64|^{3.88})$ is the mass
function for which we use the expression given by \citet{jenki}.
In this expression, $\Delta_{\mathrm{z}}=\delta_c(z)/\Delta(M)$,
in which $\delta_c(z)$ is the overdensity threshold for spherical
collapse by redshift $z$, and $\Delta(M)$ is the rms of present
variance of the fluctuations in a sphere containing a mean mass
$M$. The overdensity threshold is given by $\delta_c(z)=1.68/D(z)$
for the $\Lambda$CDM cosmology \citep{nfw97}, where
$D(z)=g[\Omega(z)]/[g(\Omega_{\mathrm{m}})(1+z)]$ is the linear
growth function of the density perturbation \citep{carroll}, in
which $g(x)=0.5x(1/70+209x/140-x^2/140+x^{4/7})^{-1}$ and
$\Omega(z)=\Omega_{\mathrm{m}}(1+z)^3
/[1-\Omega_{\mathrm{m}}+\Omega_{\mathrm{m}}(1+z)^3]$. When we
calculate the variance of the fluctuations $\Delta^2(M)$, we use
the fitting formulae for CDM power spectrum $P(k)=AkT^2(k)$ given
by \citet{eisen}, where $A$ is the amplitude normalized to
$\sigma_8=\Delta(r_{\mathrm{M}}=8h^{-1}\mathrm{Mpc})$ given by
observations. Note that the mass of an NFW halo is taken to be the
virial mass
$M_\mathrm{DM}=4\pi\delta_\mathrm{vir}\bar{\rho}r_\mathrm{vir}^3/3$,
where $\delta_\mathrm{vir}\bar{\rho}$ is the average density
within $r_\mathrm{vir}$.

The key step in working out the final results of lensing
probabilities is how to calculate the lensing cross section
$\sigma(M,z)$ in Eq. (\ref{prob1}). Since we are interested in the
lensing probabilities with image separations larger than a certain
value $\Delta\theta$ (ranging from $0\sim 10$ arcseconds, for
example) and flux density ratio less than $q_{\mathrm{r}}$, the
cross section should be defined under two conditions. The first
condition is used to define the cross section of cluster-size NFW
lenses, for which, multiple images can be produced only if
$|y|\leq y_{\mathrm{cr}}$, where $y_{\mathrm{cr}}$ is the maximum
value of $y$ when $x<0$, which is determined by $dy/dx=0$ when
$f_{\mathrm{eff}}=0$ in Eq.(\ref{lq}).  On the other hand, because
of the existence of the central point mass, theoretically, every
galaxy-size lens will always produce two images. So the first
condition fails in this case and we need the second condition to
define the cross section of galaxy-size lenses, which is the
allowed upper limit of flux density ratio of lensing images in any
lensing survey experiments. So, the flux density ratio
$q_{\mathrm{r}}$ for the two images is just the ratio of the
corresponding absolute values of magnifications \citep{schne, wu},
$q_{\mathrm{r}}=|\mu_+/\mu_-|$, where
$\mu_+[y(x)]=\left(\frac{y}{x}\frac{dy}{dx}\right)_{x>0}$ and
$\mu_-[y(x)]=\left(\frac{y}{x}\frac{dy}{dx}\right)_{x<0}$. So
$y_{\mathrm{cr}}$ is determined by
$|\mu_+(y_{\mathrm{cr}})|=q_{\mathrm{r}}|\mu_-(y_{\mathrm{cr}})|$.
The cross section for images with a separation greater than
$\Delta\theta$ and a flux density ratio less than $q_{\mathrm{r}}$
is \citep{schne}
\begin{equation}
\sigma(M,z)=\pi
r_{\mathrm{s}}^2\vartheta(M-M_{\mathrm{min}})\times \cases{
y_{\mathrm{cr}}^2,&for $\Delta\theta\leq\Delta\theta_0$; \cr
y_{\mathrm{cr}}^2-y_{\Delta\theta}^2,&for
$\Delta\theta_0\leq\Delta\theta<\Delta\theta_{y_{\mathrm{cr}}}$;\cr
0,&for $\Delta\theta\geq\Delta\theta_{y_{\mathrm{cr}}}$,\cr}
\label{cross}
\end{equation}
where $\vartheta(x)$ is a step function, and $M_{\mathrm{min}}$ is
the minimum mass of halos above which lenses can produce images
with separations greater than $\Delta\theta$. From Eq.(\ref{lq}),
an image separation for any $y$ can be expressed as
$\Delta\theta(y)=r_{\mathrm{s}}\Delta
x(y)/D_\mathrm{L}^\mathrm{A}$, where $\Delta x(y)$ is the image
separation in lens plane for a given $y$. So in Eq.(\ref{cross}),
the source position $y_{\Delta\theta}$, at which a lens produces
the image separation $\Delta\theta$, is the reverse of this
expression. And $\Delta\theta_0=\Delta\theta(0)$ is the separation
of the two images which are just on the Einstein ring;
$\Delta\theta_{y_{\mathrm{cr}}}=\Delta\theta(y_{\mathrm{cr}})$ is
the upper-limit of the separation above which the flux ratio of
the two images will be greater than $q_{\mathrm{r}}$. Note that
since $M_{\mathrm{DM}}$ ($M_{15}$) is related to $\Delta\theta$
through
$r_\mathrm{s}=(1.626/c_1)(M_{15}^{1/3}/\left[\Omega_\mathrm{m}
(1+z)^3+\Omega_{\Lambda}\right]^{1/3})h^{-1}\mathrm{Mpc}$
\citep{li}, we can formally write
$M_{\mathrm{DM}}=M_{\mathrm{DM}}(\Delta\theta(y))$, and determine
$M_{\mathrm{min}}$ for galaxy-size lenses by
$M_{\mathrm{min}}=M_{\mathrm{DM}}(\Delta\theta(y_{\mathrm{cr}}))$,
and for cluster-size lenses by
$M_{\mathrm{min}}=M_{\mathrm{DM}}(\Delta\theta(0))$. In the latter
case we have used the fact that the separation of the outermost images
is insensitive to the value of $y$ in cluster-size NFW lenses.

To compare the predicted lensing probabilities with the combined
data from JVAS/CLASS, magnification bias must be considered. For
the JVAS/CLASS sample, we use the result given by \citet{li}:
$B(M,z)=2.22A_{\mathrm{m}}^{1.1}(M,z)$, where
$A_{\mathrm{m}}(M,z)=\Delta x(y=0)/y_{\mathrm{cr}}$.


\begin{figure}
\epsscale{0.65} \plotone{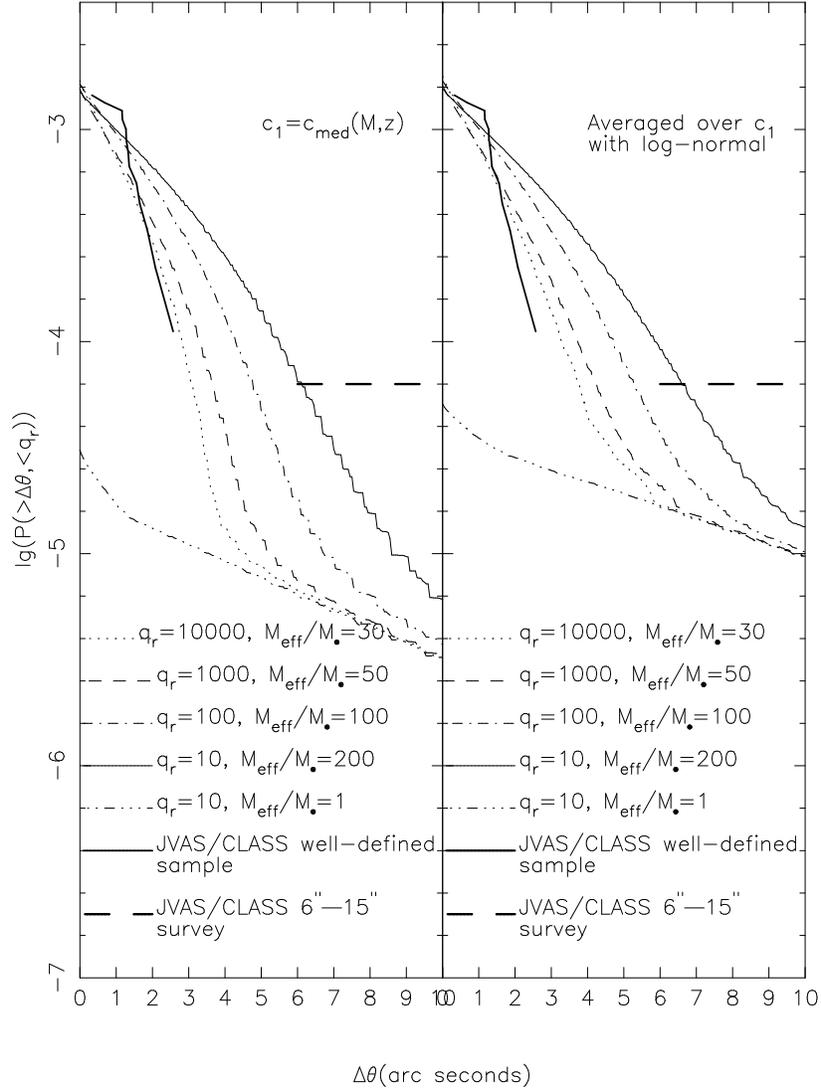}
 \caption{Predicted lensing
probability with image separations $>\Delta\theta$ and flux density
ratios $<q_{\mathrm{r}}$ in $\Lambda$CDM cosmology. The cluster-size
lens halos are modelled by the NFW profile, and galaxy-size lens halos
by NFW+BULGE. Instead of SIS, we treat the bulge as a point mass,
its value $M_{\mathrm{eff}}$ is so selected for each
$q_{\mathrm{r}}$ that the predicted lensing probability can match the
results of JVAS/CLASS represented by histogram. In the left panel,
the value of the concentration parameter $c_1$ is taken to be its
median for any given halo mass and redshift. In the right panel,
the scatter in $c_1$ is considered by averaging the probability
with the well known log-normal distribution. In both panels,
$\sigma_8=0.95$, and the null result for lenses with
$6^{''}\leq\Delta\theta\leq 15^{''}$ of JVAS/CLASS is
shown with the thick dashed horizontal line indicating the upper
limit.} \label{fig1}
\end{figure}


One major uncertainty in the estimate of $P(>\Delta\theta,
<q_{\mathrm{r}})$ by the NFW halo arises from the assignment of
the concentration parameter $c_1$ to each halo of mass $M$. There
exist several empirical fitting formulae or analytic models to
fulfill the task. However, for a given halo mass and redshift,
there is a scatter in $c_1=r_\mathrm{ vir}/r_\mathrm{ s}$ value
that is consistent with a log-normal distribution with standard
deviation $\sigma_c=\Delta(\log c)\approx 0.18$
\citep{jing,bul01}. We take into account the scatter in $c_1$ by
averaging the lensing probability with the log-normal distribution
(the right panel of figure \ref{fig1} and all the panels of figure
\ref{fig2})

Another Major uncertainty in predicting $P(>\Delta\theta,
<q_{\mathrm{r}})$ arises from the considerable uncertainty
regarding the value of the CDM power spectrum normalization
parameter $\sigma_8$, so it would be useful to consider the effect
of varying this parameter within its entire observational range, roughly
$\sigma_8=0.7\sim 1.1$ (see Fig. \ref{fig2}).

\section{Discussion and conclusions}
The lensing probabilities predicted by Eq. (\ref{prob1}) and
calculated from the combined JVAS/CLASS data are compared in
Fig.1. Since we are interested in the degeneracy between
$q_{\mathrm{r}}$ and $M_{\mathrm{eff}}$ in matching the predicted
results to observational data, we have calculated the lensing
probabilities for four different values of $q_{\mathrm{r}}$ and
their corresponding values of $M_{\mathrm{eff}}$, as indicated in
Fig. \ref{fig1}. We have assumed a ``cooling mass" of
$M_{\mathrm{c}}\approx 5\times 10^{13}h^{-1}M_{\sun}$, above which
the lenses are assigned the NFW profile and below which the lenses
are the NFW+point mass. The predicted lensing probabilities are
quite sensitive to $M_{\mathrm{c}}$, the value of $M_{\mathrm{c}}$
used here is higher than those preferred by other authors
\citep{kocha,li,sarbu}. A lower value (e.g.,
$M_{\mathrm{c}}\approx 10^{13}h^{-1}M_{\sun}$) requires a larger
$M_{\mathrm{eff}}$ for each  $q_{\mathrm{r}}$, however, this will
not affect our conclusions, since our main goal is not to
``measure" the exact value of $M_{\mathrm{eff}}$. The combined
JVAS/CLASS survey is now completed, a subset of 8,958 sources form
a well-defined statistical sample containing 13 multiply-imaged
sources (lens systems) suitable for analysis of the lens
statistics. One of the four observational selection criteria for
this ``well-defined" sample is: the image components in lens
systems must have separations $\geq 0.3$ arcsec and the ratio of
the flux densities of the brighter to the fainter component in
double-image systems must be $q_{\mathrm{r}}\leq 10$ \citep{chae}.
The observed lensing probabilities can be easily calculated:
$P_{\mathrm{obs}}(>\Delta\theta)=N(>\Delta\theta)/8958$, where
$N(>\Delta\theta)$ is the number of lenses with separation greater
than $\Delta\theta$ in 13 lenses.
$P_{\mathrm{obs}}(>\Delta\theta)$ is plotted as a histogram in
both panels of Fig. \ref{fig1}.

First of all, as shown in Fig. \ref{fig1}, when averaged over
concentration parameter $c_1$ with the log-normal distribution
(right panel), the probabilities  are increased considerably at
larger image separations and only slightly increased at smaller separations
for all cases, and the ``scatter" among the four cases is reduced.
So the scatter in $c_1$ should be considered when one uses the NFW
profile to constrain some related parameters.

It is also shown in Fig. \ref{fig1} that, for low flux ratios
($q_{\mathrm{r}}\leq 10$, as for the JVAS/CLASS survey), the NFW plus
a single SMBH model produces far too few small separation lenses
(the dash-dot-dot-dot line in the figure). This is confirmed by
the fact that, among the 22 confirmed lenses in JVAS/CLAS, none
has a fainter image very close to the center of the lens
\citep{browne02}. In other words, up to date strong gravitational
lensing effect of a single SMBH has not been found.

A larger flux ratio requires a smaller fraction of the bulge
mass as a point mass. It's interesting to note that, when
$q_{\mathrm{r}}\approx 10^4$ and $M_{\mathrm{eff}}\approx
30M_{\bullet}$, and when no scatter in $c_1$ is included (the
dotted line in the left panel of Fig. \ref{fig1}), the predicted
lensing probability fit quite well the JVAS/CLASS results. As
pointed out earlier, this conclusion can be equivalently obtained
with the model of two populations of halos, the combination of SIS
and NFW, when no constraints on flux density ratio
$q_{\mathrm{r}}$ are taken into account (i.e.,
$q_{\mathrm{r}}\sim\infty$. Note that, the flux density ratios
with values as high as $q_{\mathrm{r}}\sim 10^4$ and
infinity have approximately the same effect on the predicted
probabilities). However, since the flux density ratio for the
well-defined JVAS/CLASS sample is limited to $q_{\mathrm{r}}\leq
10$, the above mentioned good fit cannot be regarded as reflecting
the real nature of CDM halos.  In fact, to match the JVAS/CLASS
results, for $q_{\mathrm{r}}\leq 10$, about $20\%$ of the bulge
mass is required as a point mass (i.e., $M_{\mathrm{eff}}\approx
200M_{\bullet}$, the solid lines in Fig. \ref{fig1}), while for
$q_{\mathrm{r}}\leq 10^4$, only $3\%$ of bulge mass is needed.
This difference is very important when one attempts to use the
observational results of JVAS/CLASS to constrain the density
profile of galactic bulge and/or dark matter halos. It's well
known that, the flux density ratio will reduce the lensing
cross section considerably. All the models without considering this
would have overestimated the lensing probabilities. In our NFW +
Point mass model, we can adjust the value of the point mass to the
corresponding flux density ratio to match the observational
results, and will never fail. In the NFW+SIS model, however,
the contributions of SIS to lensing probabilities cannot be
changed. Put in another way, after considering the flux density
ratio limited by JVAS/CLASS, NFW+SIS model cannot reproduce the
observational results, not even marginally, if some other
parameters like $M_\mathrm{c}$ are the same as used here.
\begin{figure*}
\epsscale{0.80} \plotone{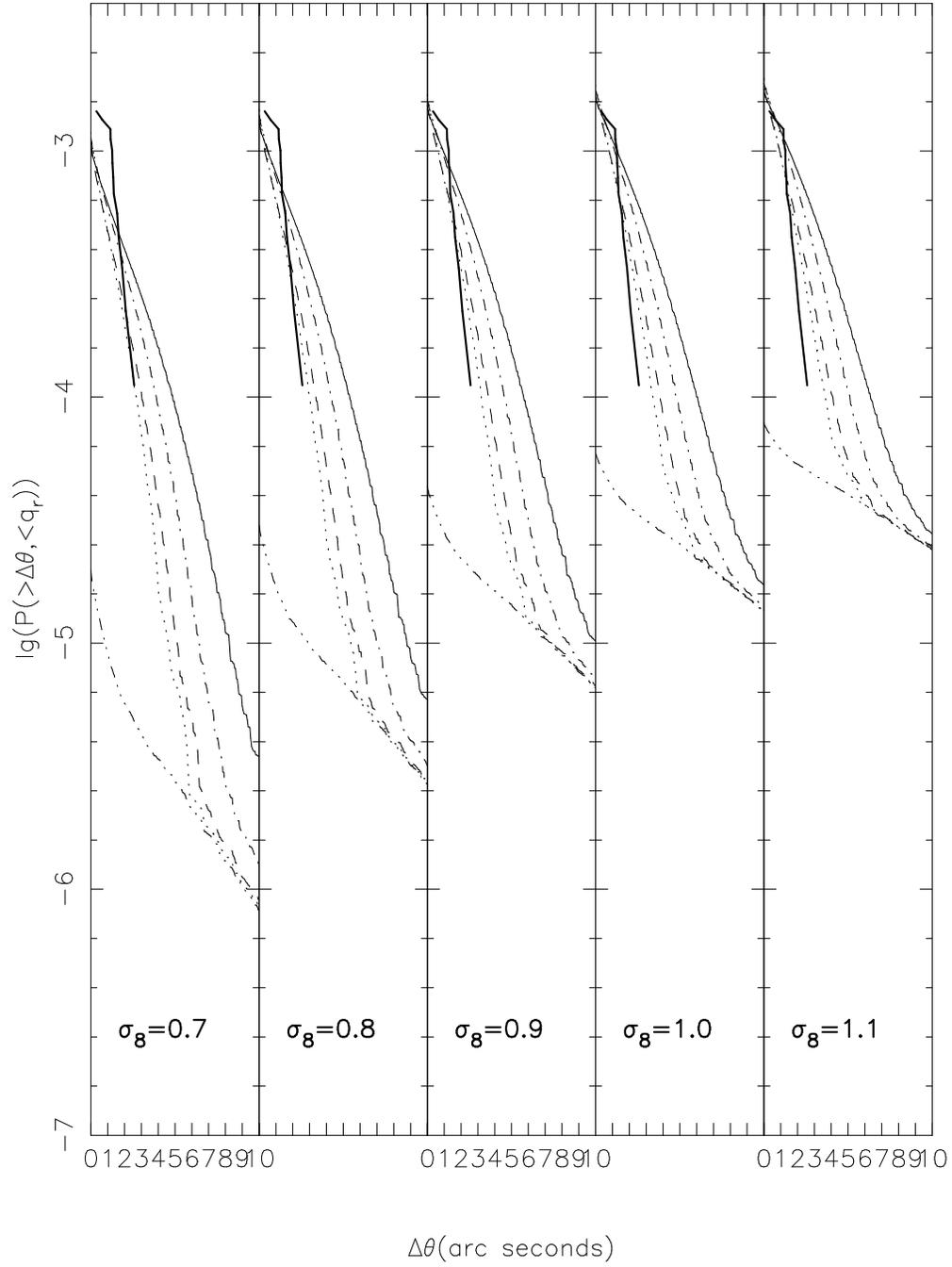}
 \caption{The same as the right panel of Fig. \ref{fig1} except
 $\sigma_8$ for each panel here. From left to right, $\sigma_8$ is
 0.7, 0.8, 0.9, 1.0 and 1.1, respectively. }\label{fig2}
\end{figure*}
We have noted that, the solid lines in both panels of Fig.
\ref{fig1} can only marginally fit the observations of JVAS/CLASS
at larger image separations. But it is still acceptable, in the
sense that, the curve is lower than the upper limit provided by
the JVAS/CLASS survey for $6^{''}\leq\Delta\theta\leq 15^{''}$
\citep{phillips}. We know that among the total of 22 confirmed

lens systems of the JVAS/CLASS survey, 21 of them have image
separations between $0.3^{''}$ and $3^{''}$, and one of them
(CLASS B2108+213) has an image separation of $4.6^{''}$
\citep{browne02}. The null result of the JVAS/CLASS survey for
$6^{''}\leq\Delta\theta\leq 15^{''}$ means that the NFW profile is
suitable for all mass sizes of halos, when a certain fraction of
the mass of each galactic bulge is considered.


Although we believe that the reasonable match to the observations
of JVAS/CLASS is $q_\mathrm{r}\leq 10$ and
$M_\mathrm{eff}/M_\bullet=200$ (the solid line in each panel of
Fig. \ref{fig1}), it is helpful if we treat the four cases (four
upper lines matching the histogram) as a whole to constrain the
sensitive parameter $\sigma_8$. We plot in each panel of Fig.
\ref{fig2} the averaged lensing probability as a function of image
separation $\Delta\theta$. All the parameters are the same as
those in the right panel of Fig. \ref{fig1} except for $\sigma_8$, for
which five different values within the entire observational range
(from 0.7 to 1.1, as explicitly indicated) have been
investigated to see their effect on the predicted lensing probabilities.
Clearly, the larger values of $\sigma_8$ will produce the higher
probabilities. In the case of $\sigma_8=0.7$, too few lenses are
produced at small image separations, so a low value of $\sigma_8$
($\leq 0.7$) is unlikely. The best fit is $\sigma_8\sim
1.0$. This result is very close to that obtained most recently by
\citet{bahcall}, in which $\sigma_8$ is determined from the
abundance of massive clusters at redshifts $z=0.5 - 0.8$. Our
 result is also in excellent agreement with that of
\citet{komatsu}($\sigma_8=1.04\pm 0.12$ at $95\%$) suggested by
the excess cosmic microwave background fluctuations detected on
small scales by the CBI \citep{mason} and the BIMA \citep{dawson}
experiments.

In summary, strong gravitational lensing probability is quite
sensitive to the cooling mass $M_\mathrm{c}$, concentration parameter
$c_1$, flux density ratio $q_\mathrm{r}$ and the spectrum
normalization parameter $\sigma_8$. Including the scatter in $c_1$
in the calculations would increase the lensing probabilities
considerably at larger image separations, while neglecting
$q_\mathrm{r}$ would overestimate the lensing probabilities. In
our NFW + point mass model, $20\%$ of the galactic bulge mass
is required as a point mass to match the JVAS/CLASS results; the
low values of $\sigma_8$ ($\leq 0.7$) are ruled out, and the preferred
value in our model is $\sigma_8\approx 1.0$.

\begin{acknowledgements}
I thank Li-Xin Li for helpful discussions and the
anonymous referee for
useful suggestions. This work was supported by the National
Natural Science Foundation of China under grant No.10233040.
\end{acknowledgements}

\end{document}